# Studies of the Temperature-Driven Flow Lines and Phase Transitions in a Two-Dimensional Si/SiGe Hole system

C.-T. Liang, C. F. Huang, Yu-Ming Cheng, Tsai-Yu Huang, Y. H. Chang, and Y. F. Chen

*Department of Physics, National Taiwan University, Taipei 106, Taiwan, R. O. C.*

We have performed low-temperature transport experiments on a Si/SiGe hole system. The meaured transverse and longitudinal conductivities $\sigma_{xy}$ and $\sigma_{xx}$ allow us to study the magnetic-field-induced transitions in the system. In particular, we present the first study of the temperature-driven flow lines in the "anomalous Hall insulator" regime near a Landau level filling factor $\nu$=1.5. The "anomalous" temperature-driven flow lines could be due to the unusual energy level scheme in a Si/SiGe hole system. Moreover, for 3<$\nu$<5, there is a temperature-independent point in $\rho_{xx}(B)$, $\rho_{xy}(B)$, $\sigma_{xx}(B)$, and $\sigma_{xy}(B)$ which corresponds to a boundary of the quantum phase transition.

## I. Introduction

Investigation of the phase transitions in the quantum Hall effect is an important research topic in the 2DHS (two-dimensional hole systems) or 2DES (two-dimensional electron system) [1-10]. According to the global phase diagram suggested by Kivelson, Lee, and Zhang, [1] a 2DES or 2DHS is an insulator when highly disordered. At low disorder, there are a series of quantum Hall states and magnetic-field-induced transitions at various magnetic field *B*. The global phase diagram also gives us the selection rule for such magnetic-field-induced phase transitions in the quantum Hall effect. For example, there are no insulators between neighboring integer quantum Hall states. The global phase diagram, however, is a zero temperature picture with infinte sample size. At a finite temperature *T*, the magnetic-field-induced transitions are governed by the scaling theory. [1,4,9-13] The width in *B* of each transition region between two quantum Hall states is proportional to $T^k$, and shrinks to a single metallic point as $T\rightarrow 0$, where *k* is the critical exponent.

Early theoretical work on the integer quantum Hall effect describes a two-parameter scaling [12] in which both the dissapative conductance $\sigma_{xx}$ and the Hall conductance $\sigma_{xy}$ vary with sample length *L*. Note that the remormalisation-group function can be illustrated by flow lines [12]. After successive length-scale transformations the flow lines are directed towards fixed points $(\sigma_{xx},\sigma_{xy})=(0,n)$ (in the units of $e^2/h$ in which $n=0,1,2,\cdots$ is the Landau level index. In addition to these "localisation fixed points" which describe the localisation of the electron wave functions near the Fermi energy, there are "intermediate-coupling fixed points" on $\sigma_{xy}=n+1/2$ which are related to transitions between quantum Hall states. Note that this theory was developed for *T*=0 and the length scale transformations are accomplished by varing the sample length *L*. It has been shown that, in practice, the effective sample size can be varied by changing the temperature [9]. The studies of the "temperature-driven flow lines" in GaAs systems and more recently in Ge/SiGe systems [11] supports the two parameter scaling theory of Pruisken [12].

In most cases, the observed quantum Hall states under unresolved spin-splitting are of even Landau level filling factors [2-4,9-11]. It was found that the critical exponent *k* depends on whether the spin-splitting is resolved or not [10]. Spin-splitting also has effects on the $\sigma_{xy}$-$\sigma_{xx}$ diagrams [9-15]. It was shown that the positions of the intermediate-coupling fixed points and the temperature-driven flow lines depend on the ratio of the spin-splitting to the Landau-level spacing [15].

Recently there has been a great deal of interest in SiGe hole gases. For example, at a Landau level filling factor ν=1.5, an insulating phase [5-9] observed in such a system is not fully understood at present. It is called a "Hall insulator" since although $\rho_{xx}$ approaches infinity, the Hall resistivity remains finite at approximately $h/2e^2$. This magneto-driven quantum Hall transition was not expected within the global phase diagram scheme of Kivelson, Lee, and Zhang. Moreover, for a Si/SiGe hole gas the observed quantum Hall states are of odd filling factors at low magnetic fields, indicating that spin-splitting is comparable with the spacing between adjacent Landau levels. This provides an interesting system for the study of magnetic-field-induced transitions in the quantum Hall effect. In this paper, we present low-temperature transport measurements on a Si/SiGe hole gas. In particular, we report the *first* study of the temperature-driven flow lines in the "Hall insulator" regime at around ν=1.5. At higher Landau level filling factors, 3<ν<5, there is a temperature-independent point in $\rho_{xx}(B)$ and $\rho_{xy}(B)$ which corresponds to a boundary of the quantum phase transition.

## II. Experimental results

The Hall bar device used in this work was made from a high-quality $Si/Si_{0.88}Ge_{0.12}$ heterostructure [8]. In our structure, holes are confined in two dimensions in a SiGe layer ~40 nm thick, sandwiched between the Si substrate and the cap layer. The carrier concentration of the 2D hole gas (2DHG) is $3.3\times10^{11}$ cm$^{-2}$ with a mobility of ~6000 cm$^2$V$^{-1}$s$^{-1}$. The experiments were performed in a top-loading He$^3$ cryostat. Four-terminal magnetoresistivities were measured using standard phase-sensitive lock-in techniques.

Figure 1 (a) shows the four-terminal longitudinal and transverse magnetoresistivity $\rho_{xx}(B)$ and $\rho_{xy}(B)$ at various temperatures $T$. At zero magnetic field, there is a small decrease in $\rho_{xx}$ as $T$ is lowered, characteristic of the metallic phase [16,17] close to the metal-insulator transition. An interesting feature is that there is a point $A1$ which is temperature-independent around a critical field $B$=3.1 T for both $\rho_{xx}$ and $\rho_{xy}$ for 3<$\nu$<5. We suggest that the temperature independent point corresponds to a quantum phase transition boundary and note that similar behaviour was observed in a higher quality SiGe hole system [18].

As shown in figure 1 (a), an anomalous Hall insulator near $\nu$=1.5 is observed. The two temperature-independent points in $\rho_{xx}$ near $B_{B1}$=8.1T and $B_{B2}$=9.8T are the boundaries of the "Hall insulator". Note that the distance in $B$ between these two critical points is the "width" of the anomalous Hall insulator regime. Therefore the transition region between the $\nu$=1 and 2 quantum Hall states *will not* shrink to a single point, which is different from the usual systems with spin-splitting much smaller than the Landau-level spacing [9-11,14]. Converting $\rho_{xx}$ and $\rho_{xy}$ into $\sigma_{xx}$ and $\sigma_{xy}$, we plot the corresponding $\sigma_{xx}(B)$ and $\sigma_{xy}(B)$ in Fig. 1(b). We can see that a large $\rho_{xx}$ centred around $\nu$=1.5 causes $\sigma_{xy}$ to drop well below its quantised value $\nu e^2/h$ for 1<$\nu$<3.

Figure 2(a) shows the $\sigma_{xy}$-$\sigma_{xx}$ diagram for 1<$\nu$<3 at $T$=0.3K. As $\nu \to 1$, $(\sigma_{xy}, \sigma_{xx}) \to (e^2/h, 0)$ because of the appearence of the quantum Hall state $\nu$=3. In addition, there is a point $Q$ at which $\sigma_{xy}(\sigma_{xx})$ crosses itself. The $T$-driven flow lines are complicated in our system thus we divide our results into four parts as shown in Fig. 2(b)-(e). It is evident that $(\sigma_{xy}, \sigma_{xx}) \to (e^2/h, 0)$ for $\nu \to 1$, as illustrated by the $T$-driven flow lines as shown in Fig. 2(b). Figure 2(c) and 2(d) show the T-driven flow diagrams near the two critical points $B_{B1}$=8.1T and $B_{B2}$=9.8T, respectively. It is evident that $(\sigma_{xy}, \sigma_{xx}) \to (0,0)$ as $T \to 0$ for the "Hall insulator" since $\rho_{xx}$ approaches infinity whereas $\rho_{xy}$ remains finite. We can clearly see that the directions of the flow lines in the "Hall insulator" are different from those near a quantum Hall state. $B_{B2}$, labelled as full squares, is, in fact, the boundary of the "Hall insulator" as well as the intermediate-coupling fixed point. It is worth mentioning that near $B_{B1}$, $\rho_{xy}$ slowly varies at different temperatures. Thus, although $B_{B1}$ corresponds to the boundary of the "Hall insulator" regime, the intermediate-coupling fixed point is estimated to be at around $B$=8.2T, ~0.1T higher than $B_{B1}$. To the right of this fixed point, the flow lines all show a "kink". In contrast, to the left of the fixed point, the lines flow directly (0,0) as $T \to 0$. The main finding is that, in contrast to a *single* intermediate-coupling fixed point between two neighoring quantum Hall states previously predicted and observed [9,12], there are *two* intermediate-coupling points in a SiGe hole system. The reason for this is the existence of the "anomalous Hall insulator" near $\nu =1.5$. Our new experimental results on *T*-driven flow lines, together with the poineering work on the "Hall insulator" [5-9] urge further investigations in order to understand the physical origin of this "anomalous insulating state". Figure 2 (e) shows the *T*-driven flow lines for $2<\nu<3$. At around *A*2, the directions of the *T*-driven flow lines seem ambiguous. At present, the physical origin of this effect is not fully understood and awaits further experimental and theoretical investigations.

Finally we turn our attention to low-field results. Figure 3 shows experimental results for $\sigma_{xy}(T)$ and $\sigma_{xx}(T)$ plotted as *T*-driven flow lines from *T*=0.3 to 1.648K. Each line corresponds to a fixed *B*. As shown in Fig. 3, we can see an intermediate-coupling fixed point labelled as squares. In fact, this intermediate-coupling fixed point shown in Fig. 3 corresponds to A1 (temperature-independent $\sigma_{xy}(T)$ and $\sigma_{xx}(T)$ as shown in Fig. 1(a) and (b). To the left of this point A1, which corresponds to a higher magnetic field, the flow lines are directed toward $(0,4.5e^2/h)$, since the Hall conductance plateau at $\nu =5$ is not well-quantised. In contrast to the semicircle-like feature seen previously [9,11], to the left of this intermediate-coupling fixed point, the *T*-driven flow lines are directed towards $\sigma_{xy}=2.0e^2/h$ and eventually converge at $\sigma_{xy}=3.0e^2/h$. This effect is believed to be due to strong Landau level mixing as described by the model of Ando [19]. The intermediate-coupling fixed point is close to $\sigma_{xy}=3.25e^2/h$, not the theoretical value of $\sigma_{xy}=3.5e^2/h$ [12]. We note that early experimental results on a GaAs electron system [9] did show a similar behaviour.

### III. Conclusions

In conclusion, we have presented low-temperature magnetotransport results on a Si/SiGe hole system. For Landau level filling factors $3<\nu<5$, there is a temeperature-independent point in $\rho_{xx}$ and $\rho_{xy}$, which corresponds to a boundary of a quantum phase transition. Due to the existence of the "Hall insulator" around $\nu =1.5$, the temperature-driven flow line study shows that there are *two* intermediate-coupling fixed points, in contrast to the single point previously predicted and observed in a GaAs sample. We speculate that the observed unusual temperature-driven flow lines could be due to the unusual energy level scheme in which the Landau level splitting is comparable to the Zeeman splitting in a Si/SiGe hole system. Our new results, together with previous work on the "Hall insulator" at $\nu =1.5$, urge further theoretical and experimental investigations to understand the underlying physics of this "anomalous insulating state" observed in a Si/SiGe hole gas.

**Acknowledgments**

This work was funded by the NSC, Taiwan (grant no NSC 89-2112-M-002-052 and 89-2112-M-002-084). The low-temperature measurements were performed at the high-field facilities at the Centre for Condensed Matter Sciences (CCMS), National Taiwan University (NTU). We would like to thank R. B. Dunford and F. F. Fang for fruitful discussions, T. J. Chuang at the CCMS for support, and H. H. Cheng for invaluable experimental assistance. C.T.L. acknowledges financial support from NTU.

**Figure captions**

Fig 1 (a). Four-terminal resistivity measurements of $\rho_{xx}(B)$ and $\rho_{xy}(B)$ at $T$=0.30, 0.60, 0.97, and 1.2K. A1, B1, and B2 are temperature-independent points in $\rho_{xx}(B)$. Near A2, has a weak temperature dependence. (b) Converted conductivity $\sigma_{xx}(B)$ and $\sigma_{xy}(B)$ at various temperatures $T$=0.30, 0.60, 0.97, and 1.2K.

Fig 2 (a) $\sigma_{xx}(\sigma_{xy})$ at T=0.3K. There is a point Q where $\sigma_{xx}(\sigma_{xy})$ crosses iteself. (b)-(e) Temperature-driven flow lines between $1<\nu<3$. The arrows indicate the directions of the flow lines (from high-$T$ to low-$T$) and the full squares correspond to intermediate-coupling fixed points. Note that the dotted line represents $\sigma_{xx}(\sigma_{xy})$ at $T$=0.3K.

Fig 3 Temperature-driven flow lines between $3<\nu<5$. The arrows indicate the directions of the flow lines (from high-$T$ to low-$T$) and the full squares correspond to an intermediate-coupling fixed point. The dotted line represents $\sigma_{xx}(\sigma_{xy})$ at $T$=0.3K.

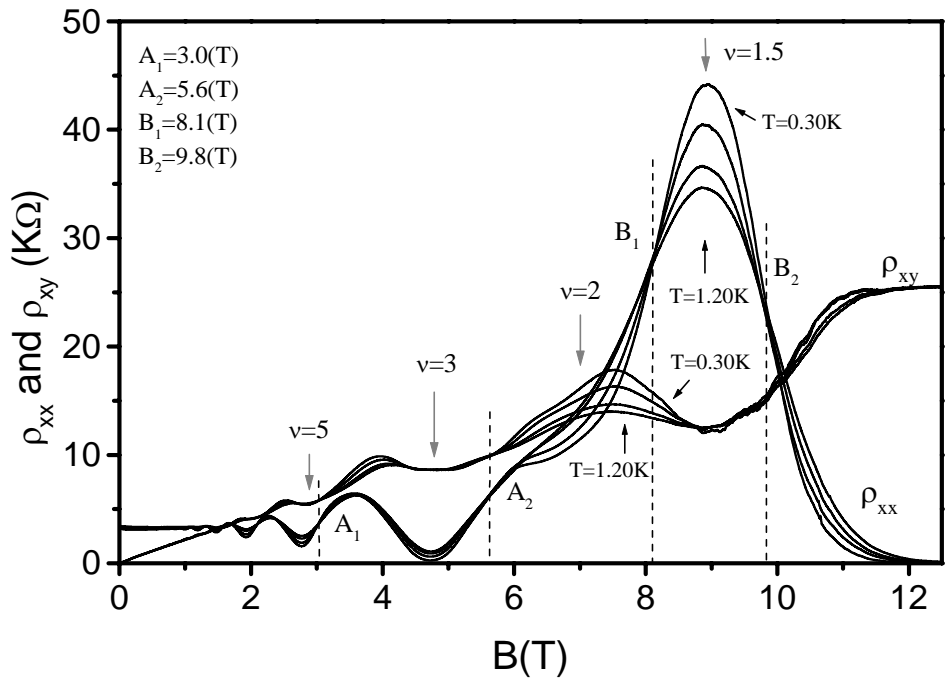

Fig. 1 (a)

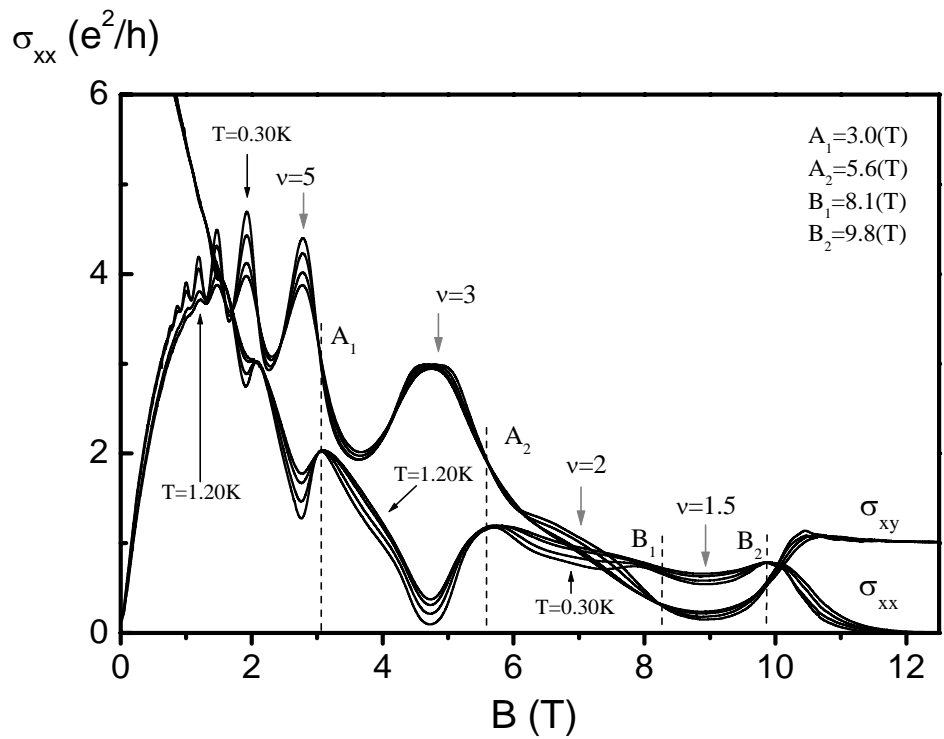

Fig. 1 (b)

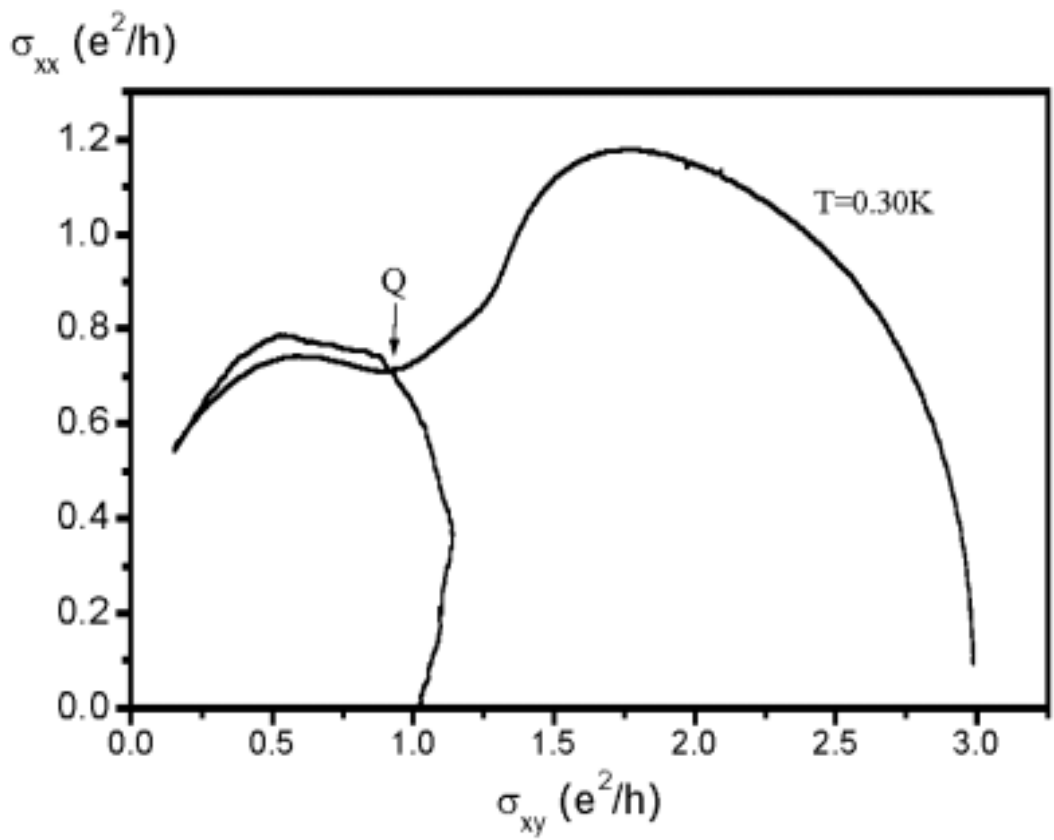

Fig. 2(a)

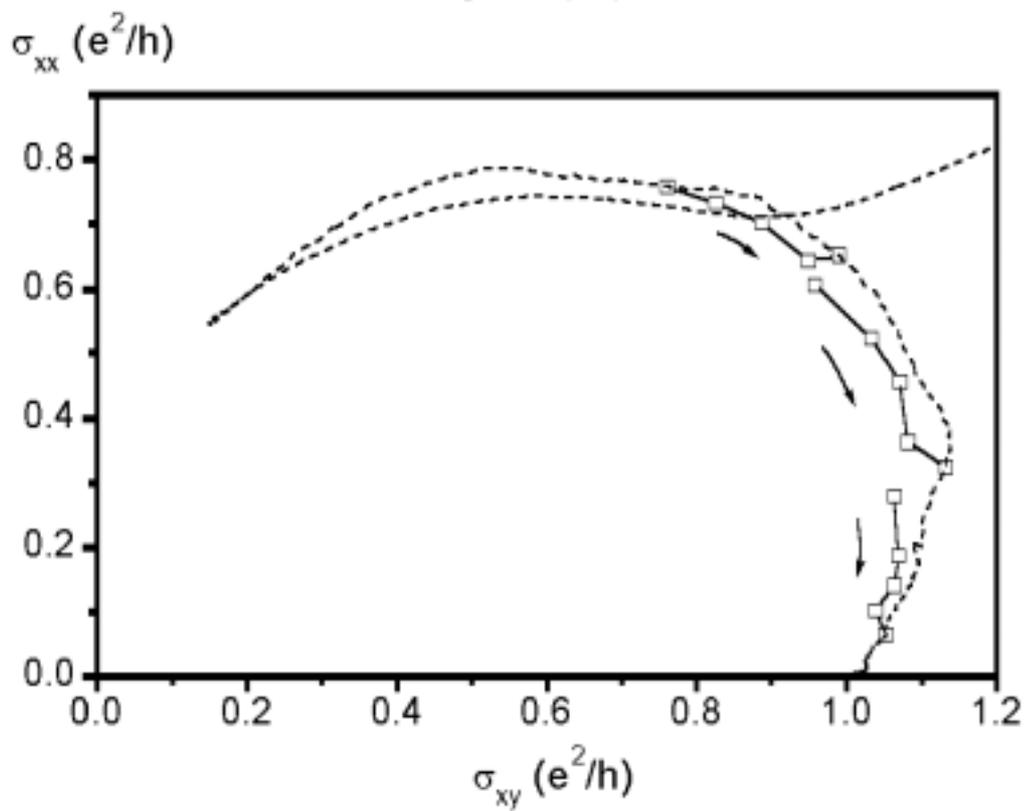

Fig. 2(b)

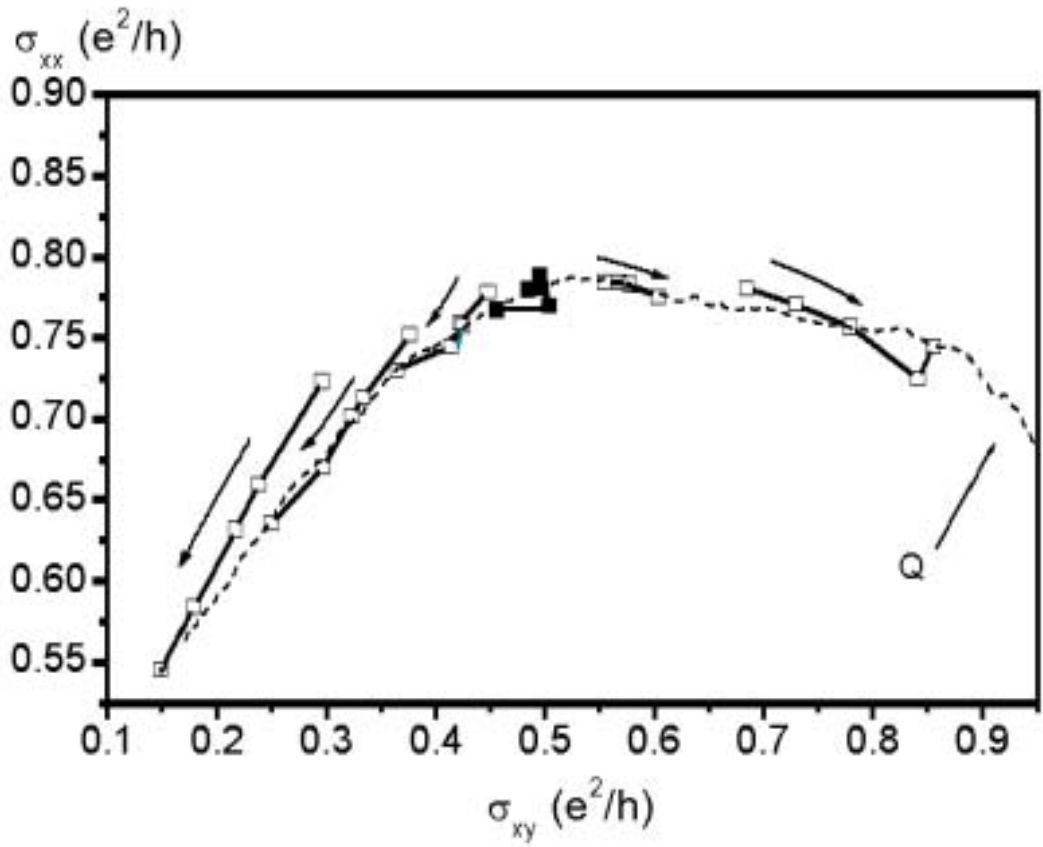

Fig. 2 (c)

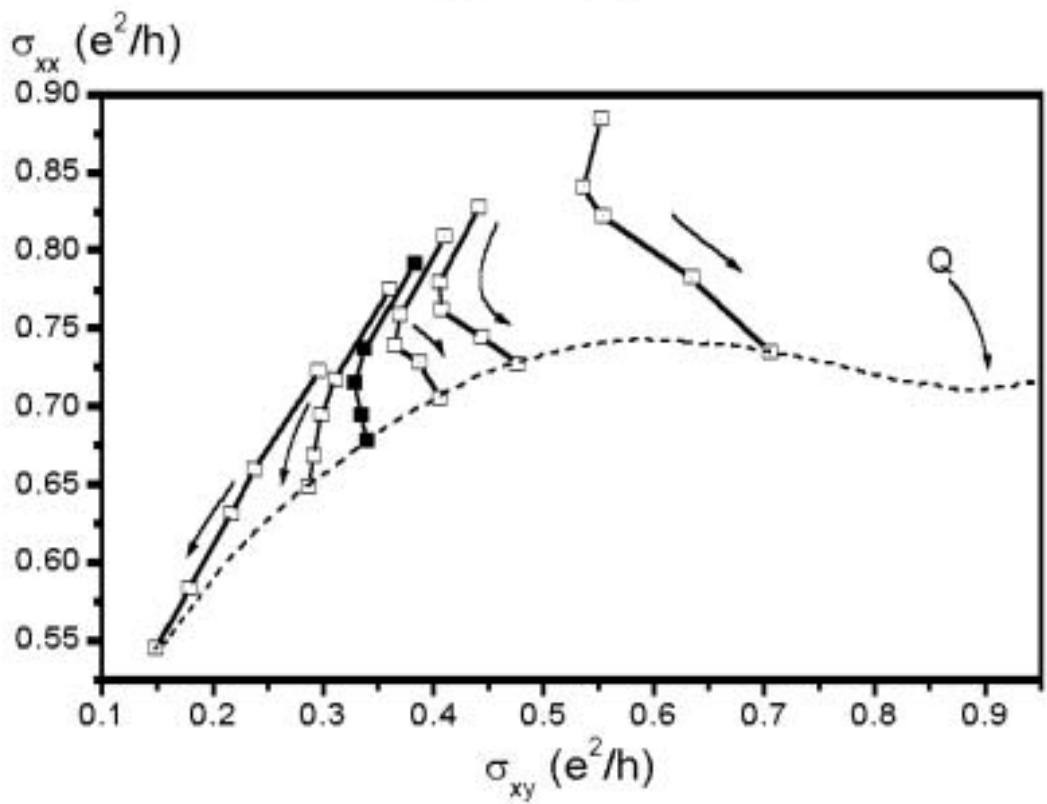

Fig. 2 (d)

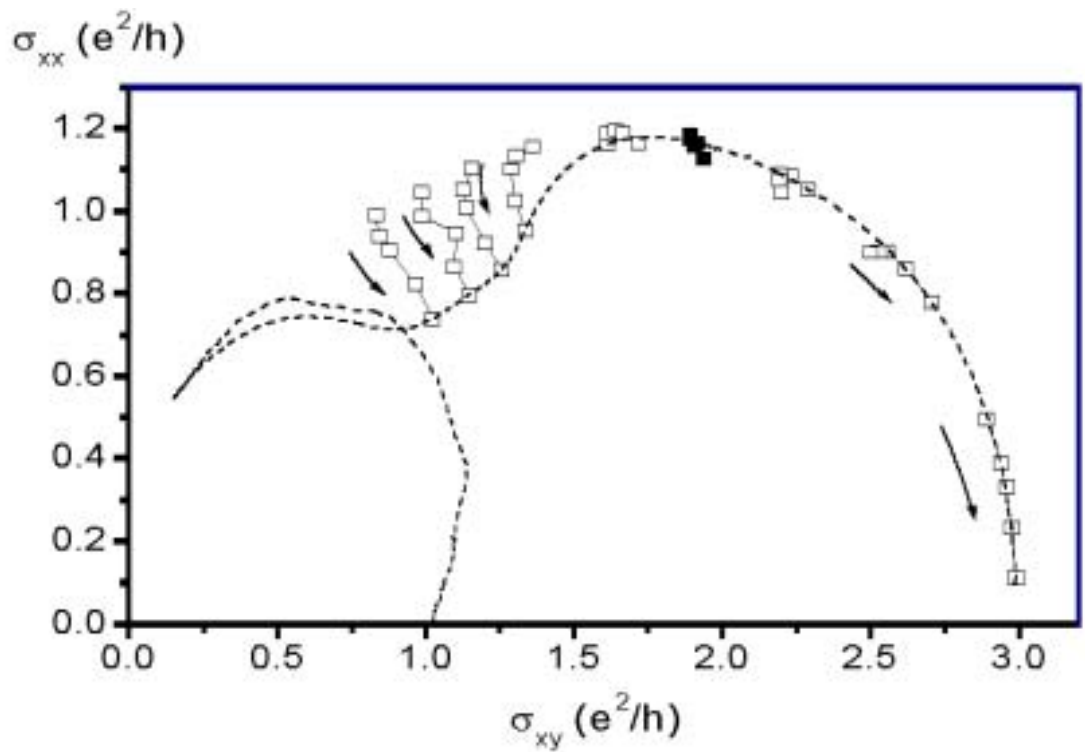

Fig. 2(e)

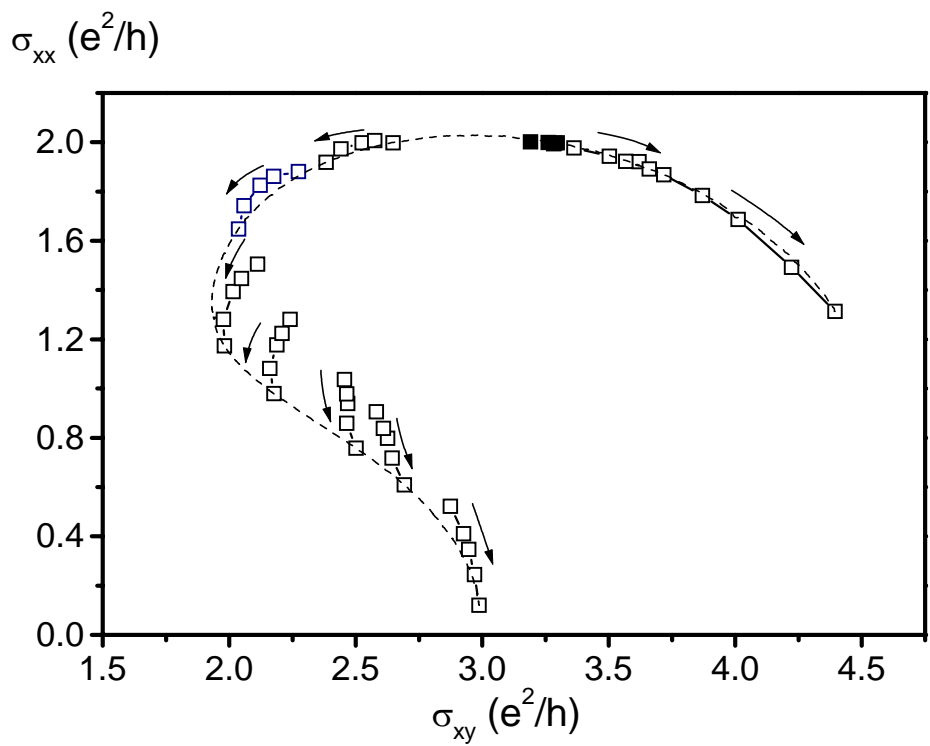

Fig. 3